\documentclass[
 aip,
 amsmath,
 amssymb,
 reprint,
]{revtex4-1}

\usepackage{graphicx}
\usepackage{dcolumn}
\usepackage{bm}

\usepackage[utf8]{inputenc}
\usepackage[T1]{fontenc}
\usepackage{mathptmx}
\usepackage{etoolbox}
\usepackage[colorlinks]{hyperref}
\hypersetup{
    colorlinks=true,     
    linkcolor=blue,      
    citecolor=blue,      
    filecolor=magenta,   
    urlcolor=blue,       
}

\makeatletter
\def\@email#1#2{%
 \endgroup
 \patchcmd{\titleblock@produce}
  {\frontmatter@RRAPformat}
  {\frontmatter@RRAPformat{\produce@RRAP{*#1\href{mailto:#2}{#2}}}\frontmatter@RRAPformat}
  {}{}
}%
\makeatother
\begin{document}


\title{Data-driven assessment of optimal spatiotemporal resolutions for information extraction in noisy time series data} 

\author{Domiziano Doria}
 \affiliation{%
Department of Applied Science and Technology, Politecnico di Torino, Torino 10129, Italy
}
\author{Simone Martino}
\affiliation{%
Department of Applied Science and Technology, Politecnico di Torino, Torino 10129, Italy
}
\author{Matteo Becchi}
\affiliation{%
Department of Applied Science and Technology, Politecnico di Torino, Torino 10129, Italy
}
\author{Giovanni M. Pavan}
\email{giovanni.pavan@polito.it}
\affiliation{%
Department of Applied Science and Technology, Politecnico di Torino, Torino 10129, Italy
}%

\date{\today}

\begin{abstract}
In general, comprehension of any type of complex system depends on the resolution used to examine the phenomena occurring within it. However, identifying \textit{a priori}, for example, the best time frequencies/scales to study a certain system over-time, or the spatial distances at which correlations, symmetries, and fluctuations are, most often non-trivial. Here we describe an unsupervised approach that, starting solely from the data of a system, allows learning the characteristic length scales of the dominant key events/processes and the optimal spatiotemporal resolutions to characterize them. We tested this approach on time series data obtained from simulation or experimental trajectories of various example many-body complex systems ranging from the atomic to the macroscopic scale and having diverse internal dynamic complexities. Our method automatically analyzes the system data by analyzing correlations at all relevant inter-particle distances and at all possible inter-frame intervals in which their time series can be subdivided, namely, at all space and time resolutions. The optimal spatiotemporal resolution for studying a certain system thus maximizes information extraction and classification from the system’s data, which we prove to be related to the characteristic spatiotemporal length scales of the local/collective physical events dominating it. This approach is broadly applicable and can be used to optimize the study of different types of data (static distributions, time series, or signals). The concept of “optimal resolution” has a general character and provides a robust basis for characterizing any type of system based on its data, as well as to guide data analysis in general.
\end{abstract}

\pacs{}

\maketitle

\section*{Introduction}

Understanding the physics of complex many-body systems with rich internal dynamics is typically a challenging task~\cite{ten_wolde_enhancement_1997, baletto_structural_2019, gasparotto_identifying_2020, bochicchio_how_2019, de_marco_controlling_2021, cioni_innate_2023, crippa_detecting_2023}. This holds true across various scales~\cite{cho_dynamics_2021}, from complex phenomena at the molecular level, such as nucleation events, phase transitions and ensemble dynamic adaptations in complex molecular systems~\cite{ten_wolde_enhancement_1997}, to non-trivial collective behaviors in complex scale systems such as fish schools~\cite{porfiri_inferring_2018, butail_model-free_2016}, bird flocks~\cite{cavagna_scale-free_2010, attanasi_information_2014, nagy_hierarchical_2010} and socioeconomic or stock market systems~\cite{liu_quantifying_2021, mantegna_introduction_1999, duan_network_2022}. 
Typically, the first challenge in studying such complex systems is to extract relevant information about the natural dynamical behavior of their constituent units, a fundamental step in reconstructing the complex internal physics that characterize them~\cite{martino2024data}. 

The study of a complex system typically begins with the acquisition of trajectories, such as the positions and velocities of the individual units composing the system, obtained either through experiments or simulations. An efficient method is required to extract physically relevant information from these trajectories. A common approach involves using descriptors that convert raw trajectories into time series that can represent functions of positions (e.g. mutual arrangements) and / or velocities of the constituent units over time~\cite{kathirgamanathan_feature_2020, schmidt_human-based_2023}. 
The choice of descriptor is critical, as different descriptors enable the extraction of different types of physical information about the system's behavior~\cite{caruso_timesoap_2023, crippa_molecular_2022, martino2024data}. Among the various available descriptors, a distinction can be made between physics-inspired descriptors and "abstract" or general ones. Physics-inspired descriptors are tailored to study specific systems by leveraging the prior knowledge of their physics. Although they can be highly effective for investigating particular types of systems, they are typically less transferable to others. Furthermore, these descriptors may suffer from inherent biases and potentially overlook important and unforeseen information. To address these limitations, more general ``abstract" descriptors -- such as correlation functions or relative distances, to name a few commonly used examples -- can be employed. 

More advanced and complex abstract descriptors, such as the Smooth Overlap of Atomic Positions (SOAP)~\cite{bartok_representing_2013} and Atomic Cluster Expansions (ACE)~\cite{drautz_atomic_2019} have also proven to be highly useful. Unlike system-specific descriptors, these general approaches are not tailored ad hoc to any particular system, making them applicable to virtually any system type. Their unbiased nature, stemming from the lack of reliance on prior knowledge of the system's physics, allows them to be broadly applicable and transferable for studying a wide range of complex molecular systems. 
Pattern recognition approaches for analyzing datasets, such as those from ACE or SOAP, have successfully provided valuable insights into the physics of systems across scales, from molecular to macroscopic, including aqueous, metallic, and other complex systems~\cite{fitzner2019ice, Capelli2022Ephemeral, lionello2022supramolecular, cioni_innate_2023, donkor_machine-learning_2023, cioni_innate_2023, crippa_machine_2023, cioni2024sampling, donkor_beyond_2024, martino2024data, perrone2024unsupervised,cheng_ab_2019,rapetti_machine_2023}, and large-scale phenomena such as wave propagation~\cite{becchi_layer-by-layer_2024, caruso2024classification}. 
Data-mining approaches often rely on identifying high-density peaks (dominant microscopic structural motifs) in datasets, treating the data as an ensemble in which time correlations are ignored. This can potentially lead to loss of information~\cite{martino2024data, lionello2024relevant}. 

Conversely, it has been demonstrated that studying data, acquired over time series (thus accounting for their time-correlation) allows decouple temporal from spatial information, which may provide more rich information than typical pattern-recognition approaches, including details on rare events, local or collective events that originate and amplify dynamically over time, etc~\cite{caruso_timesoap_2023, crippa_detecting_2023, butler_change_2024, crippa_machine_2023, becchi_layer-by-layer_2024, cioni2024sampling, perrone2024unsupervised, martino2024data, caruso2024classification, lionello2024relevant, husic_markov_2018}. 
For example, descriptors such as timeSOAP~\cite{caruso_timesoap_2023} or Local Environments and Neighbors Shuffling (LENS)~\cite{crippa_detecting_2023} track how much and in what way the environment surrounding each unit in the system changes over time. These approaches shift the focus to identifying relevant fluctuations in noisy time series. An efficient method for this is Onion Clustering, an unsupervised and parameter-free algorithm that systematically detects statistically significant fluctuations in noisy time series of any type. It classifies these fluctuations based on their similarity, determining the number of distinct microscopic domains (clusters) can be identified in a time series as a function of the length of the signal intervals used in the analysis. 
Recent results from Onion clustering have revealed non-trivial dependencies on the temporal resolution used in the analysis. These findings show that, for a given system studied with a specific descriptor, there exists an optimal temporal resolution that maximizes the information extracted from its time series data~\cite{becchi_layer-by-layer_2024, martino2024data, lionello2024relevant}. 
The considerations for temporal resolution also apply to spatial resolution. More broadly, this raises an important question: what is the optimal resolution -- both spatial and temporal -- for capturing the physical behavior of a system? For instance, it is known that adjusting the cutoff for radial-based local descriptors (e.g., capturing different ranges of short- or long-range effects) can significantly impact analysis results. Similar to the findings from Onion Clustering regarding temporal resolution, is there an optimal spatial resolution for extracting meaningful physical information from trajectory data? Moreover, is there a reliable method to determine this optimal resolution for a given system? 

In this work, we address this fundamental problem by studying various prototypical systems, ranging from atomic and molecular scales to microscopic and macroscopic scales, which exhibit different complex internal dynamics. We systematically analyzed the information contained in the time series data of their constituent units as a function of the spatial and temporal resolution used in the analyses. We demonstrated that different types of systems have distinct optimal spatiotemporal resolutions that maximize the extraction of relevant information from their trajectories, depending on the local or collective nature of the physical events/processes that occur within them. 
The characteristic spatiotemporal length scales of the dominant physical events dominating the system are in this way revealed in a purely data-driven way, and without prior assumptions on the best resolutions to capture and characterize them. 

While identifying the optimal time and space resolutions to study a system in which little is known \textit{a priori} is challenging, this work demonstrates how such information is intrinsically present in the system's data and it provides an efficient way to extract it. 
From a fundamental point of view, we provide clear demonstrations of the impact of such a data driven approach in revealing the physics of complex systems (or phenomena) without relying on prior knowledge or preconceived assumptions. At the same time, because every set of data has its own optimal spatiotemporal resolutions that allow for maximizing information extraction, we obtain a general concept and an useful approach to guide and assist the analysis of virtually any type of data. 

\section*{Results}
\subsection*{A first case study: ice/water coexistence in dynamical equilibrium}

\begin{figure*}[ht!]
    \centering
    \includegraphics[width=.9\textwidth]{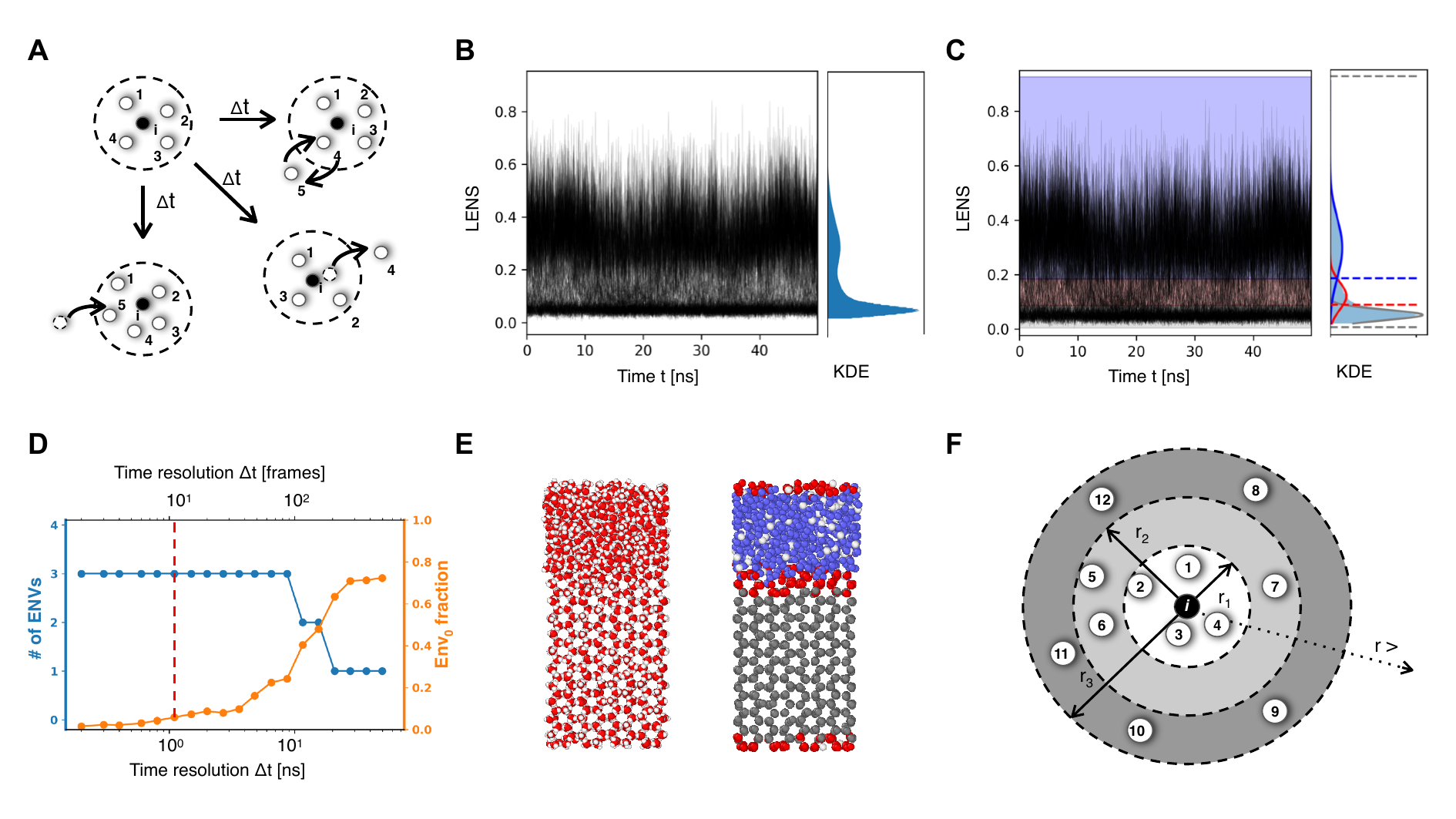}
    \caption{\textbf{Extracting information from, \textit{e.g.}, LENS time series of ice/water dynamic coexistence simulation trajectories}. 
    \textbf{A} Scheme of example local dynamical events captured by the LENS descriptor (permutation, addition, or subtraction of neighbors). 
    \textbf{B} LENS signals for all the water molecules (their oxygen atoms) in the system as a function of simulation time, and their cumulative distribution (KDE: on the right, in blue). 
    \textbf{C} Same LENS time series data with the background colored based on the three main micro-clusters detected by Onion Clustering (using a time-resolution of $\Delta t=1.1$ ns). The gaussian LENS environments are shown in the KDE (right) as solid gaussian curves, the inter-cluster thresholds are indicated as dotted horizontal lines. 
    \textbf{D} Output Onion plot. The curve in blue (primary $y$-axis) shows the number of clusters classifiable by Onion Clustering as a function of the time resolution $\Delta t$. The curve in orange (secondary $y$-axis) shows the fraction of unclassifiable data points (stored into a cluster named ENV0) as a function of $\Delta t$ (i.e., fraction of dynamical events occurring faster than the resolution of the analysis). The vertical red dashed line indicates the time-resolution $\Delta t = 1.1$~ns used for the other panels (clusters and thresholds in panel \textbf{C}). 
    \textbf{E} MD snapshot of the ice/water coexistence simulation showing the TIP4P/ICE molecules (left) and coloring them based on the clusters classified at the example time-resolution, which correspond to the bulk of ice (in gray), to the bulk of the liquid phase (in blue), and to the ice/water interface (in red). In white are the unclassifiable points: these are domains in the liquid which freeze and re-melt faster than $\Delta t=1.1$ ns~\cite{becchi_layer-by-layer_2024, Capelli2022Ephemeral, crippa_detecting_2023, caruso_timesoap_2023}. 
    \textbf{F} Schematic representation of the solvation shells accounting for the first, second, third, etc., neighbor particles around a unit {\it i} in a generic system: using different cutoff radii means capturing different types of events (e.g., local {\it vs.} non-local) whose relevance depends on the physics of system and is often not clear \textit{a priori}. }
    \label{fig:fig1}
\end{figure*}

As a representative case study, we focused on a prototypical molecular system with non-trivial internal dynamics. Specifically, we analyzed the atomistic Molecular Dynamics (MD) trajectories of a system consisting of 2048 TIP4P/ICE~\cite{abascal2005potential} molecules at the melting temperature~\cite{caruso_timesoap_2023}. Initially, half of the molecules are in the liquid phase, while the other half form solid ice. After equilibration, a $50$~ns production run was performed under $NPT$ conditions (see Methods section for complete details on the molecular model and simulation setup). 

Recent studies have demonstrated that a general yet relatively simple descriptor, LENS, can effectively capture the intrinsic dynamics of such system with remarkable sensitivity. As illustrated in Fig.~\ref{fig:fig1}A, the LENS value for every particle in the system was computed for each sampling interval (of length $\delta t$) along the MD trajectory. This is done by measuring the variation, frame by frame, in the string of the IDs of the neighbors within a specified cutoff radius (dashed circles in Fig.~\ref{fig:fig1}A). LENS captures events such as the addition, removal, or permutation of neighbors during $\delta t$, offering insights into the local dynamics of the environment surrounding each particle in the system. 

In this case, 2048 LENS time series were extracted -- one for each molecule, considering only the oxygen atoms -- each consisting of 500 frames ($\delta t = 100$~ps); see Fig.~\ref{fig:fig1}B. The probability kernel density estimate (KDE) of the data points revealed two distinct peaks (Fig.~\ref{fig:fig1}B, shown in blue). The first and most prominent peak occurs at LENS intensity $\sim0.05$, corresponding to molecules that are nearly static. These represent the molecules in the ice phase, vibrating around their crystalline positions. The second, lower-density peak appears at a LENS intensity of $\sim0.35$, corresponding to the more dynamic molecules in the liquid phase. 

These are the two main clusters that standard pattern recognition approaches, which analyze the entire dataset while ignoring time correlations, are readily identifed from the trajectories of this system~\cite{crippa_machine_2023, caruso_timesoap_2023}. However, recent studies have shown that additional, highly relevant information is hidden within these data, masked by the noise of these two dominant clusters~\cite{crippa_detecting_2023}. This information can be extracted by accounting for time correlations in the data of individual molecules, treating these data as time series, and systematically distinguishing meaningful fluctuations from noise, using single-point time series clustering algorithms, such as Onion Clustering~\cite{becchi_layer-by-layer_2024, martino2024data, lionello2024relevant}. 

As shown in Fig.~\ref{fig:fig1}C, Onion Clustering performs a single-point time series analysis on the trajectories of all molecules in the system, classifying them based on the similarity of their vibrational dynamics. This method employs an iterative detect-classify-archive approach with a resolution that adaptively increases at each iteration. 
Onion Clustering systematically repeats the clustering analysis across all possible temporal resolutions $\Delta t$, starting from the maximum resolution, which is twice the minimum time step (in this case, $200$~ps), and progressively reducing the resolution down to the minimum, corresponding to the total duration of the simulation ($50$~ns). 

The plot in Fig.~\ref{fig:fig1}D (blue curve) shows the number of statistically distinct environments identified as a function of the time resolution. The analysis reveals not just two but up to three distinct environments from the LENS time series. Physically, these correspond to the ice (gray in Fig.~\ref{fig:fig1}E), liquid (blue), and the ice-water interface (red). 
From the Onion plot, it is evident that these three clusters are distinguishable at resolutions of $\Delta t \leq 10$~ns. Beyond this threshold, the resolution becomes insufficient to differentiate the ice-water interface as a dynamically distinct environment, resulting in a reduction from three clusters to two. This finding suggests that the characteristic residence time of water molecules in the interface is approximately 10~ns; using a lower resolution than this prevents the identification of the interface environment as distinct. 

Notably, for $\Delta t > 20$~ns, only a single stable environment, corresponding to the ice bulk, can be identified. This indicated that some molecules within the ice bulk remain in this environment throughout the entire duration of the simulation. At these resolutions, $\Delta t>20$~ns, the analysis effectively detects one stable environment (ice), while all other data points are classified as part of an ``unclassifiable data" cluster. These unclassified molecules, at this resolution, cannot be attributed to any stable environment, reflecting the inability of such a coarse temporal resolution to resolve their dynamic behavior. 
Notice that, as the value of $\Delta t$ increases, the fraction of unclassified data points increases as well (orange curve in Fig.~\ref{fig:fig1}D), until most of the data are in this category, because the time resolution is too coarse to resolve most of the relevant system dynamics. 

For more details regarding the descriptor and the clustering method, we refer the interested reader to Refs.~\cite{crippa_detecting_2023} and~\cite{becchi_layer-by-layer_2024}. A crucial strength of Onion Clustering lies in its ability to transform a potential limitation of clustering analysis -- the dependence on the choice of time resolution -- into an advantage. By eliminating the need for the user to select a specific time resolution \textit{a priori}, Onion Clustering identifies the optimal resolution windows for studying the system's physics. In this case, the resolution of $\Delta t<10$~ns is ideal, because it allows the detection of three distinct clusters while minimizing the fraction of unclassifiable data points. 

It is worth highlighting that by repeating the analysis across various temporal resolutions and visualizing the outcomes, the approach becomes independent of any preconceived assumptions concerning the optimal temporal resolution for studying the system's physics. Nonetheless, one key parameter still influences the analysis: the cutoff used in calculating the descriptor (LENS in this case, though this applies to many other descriptors as well). It is well known that varying the cutoff when using a descriptor can significantly impact the results. This raises an important question: what is the optimal spatial resolution (i.e., the best cutoff radius) for maximizing the extraction and classification of information from the data and ultimately the understanding of the system's physics? While the answer to this fundamental question remains often elusive, the methodology presented here provides a robust framework to identify the ``optimal spatiotemporal resolution" for analyzing a system solely based on the data extracted from its trajectories.  

\begin{figure*}[ht!]
    \centering
    \includegraphics[width=.9\textwidth]{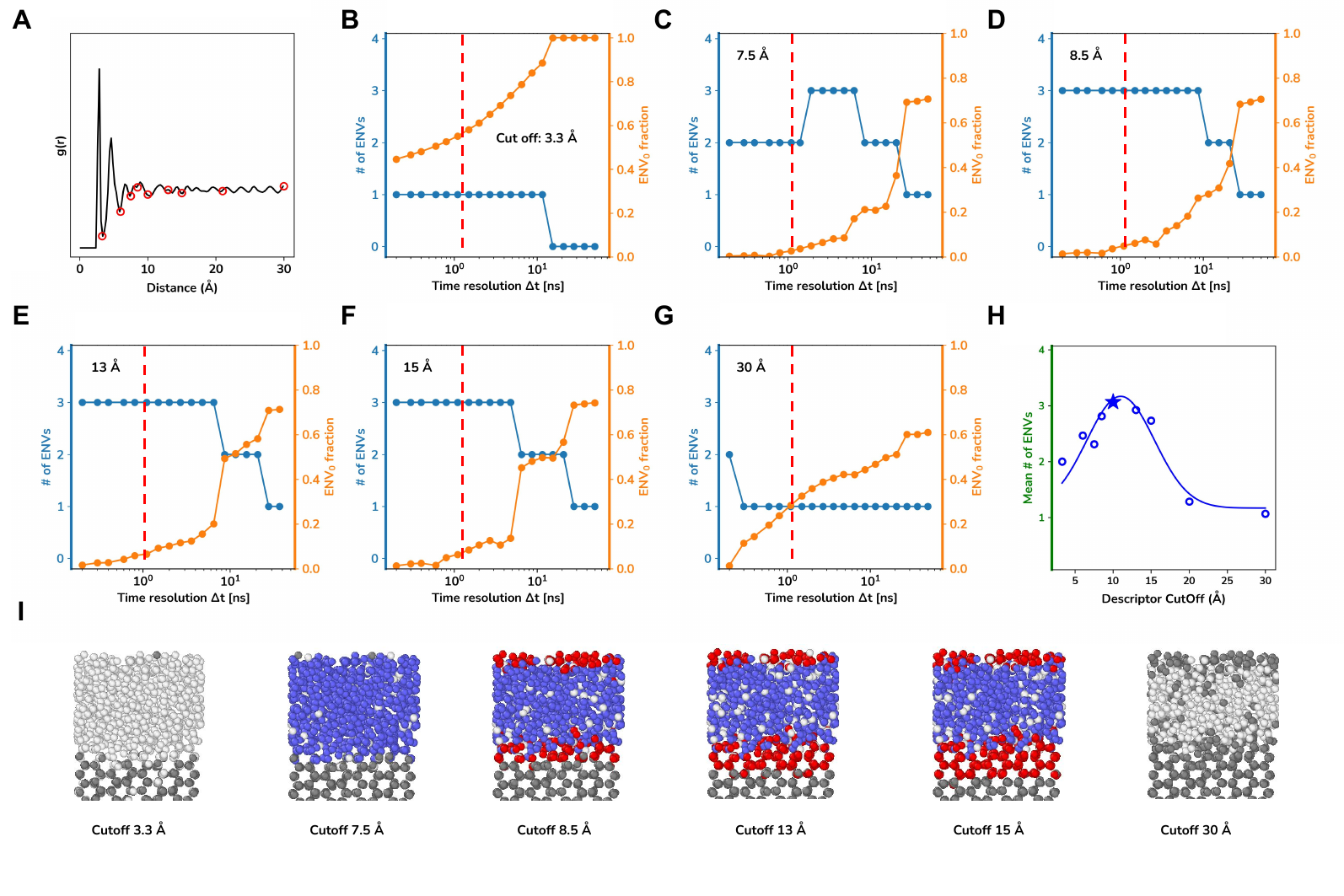}
    \caption{\textbf{Effect of spatial and temporal resolutions on information extraction from LENS time series data.} 
    \textbf{A} Radial distribution function $g(r)$ of the oxygen atoms of all water molecules in the system: relevant $g(r)$ minima are detected (red circles) and used as critical cutoff radii $r_c$ to calculate LENS signals retaining information on local \textit{vs.} non-local events/phenomena. 
    \textbf{B-G} Onion plots obtained from the analyses of the LENS time series with different $r_c$. The number of resolved micro-clusters is shown in blue, while the fraction of unclassifiable data (in the ENV0 cluster) is shown in orange. The red vertical dashed line shows the example time resolution of $\Delta t=1.1$ ns used for the snapshots in panel \textbf{I}. 
    \textbf{H} In blue: mean number of classifiable micro-clusters (ENVs) before the fraction of unclassifiable data reaches 50\%, as a function of the $r_c$ used in the analysis. The data show a clear trend (blue curve), where the maximum efficiency in information extraction-and-discretization is obtained at $r_c \sim 10$ \AA: namely, when accounting for events involving up to the 3rd-4th neighbors shell. 
    \textbf{I} Representative MD snapshot of the system where the water molecules are colored according to the clustering obtained with the different $r_c$ in the analyses of panels \textbf{B-G} at the example time-resolution of $\Delta t=1.1$ ns: the red ice/water interface can be resolved in the spatial resolution range of $8.5 \leq r_c \leq 15$ \AA, and for temporal resolutions higher than $\Delta t < 10$ ns.}
    \label{fig:fig2}
\end{figure*}

\subsection*{Optimal spatiotemporal resolution(s)}

It is well-known that complex systems can be governed by local or collective events~\cite{cioni_innate_2023,crippa_detecting_2023,caruso_timesoap_2023,de_marco_controlling_2021,butail_model-free_2016,offei-danso_high-dimensional_2022} that may emerge and amplify within them. When one does not have prior knowledge of a system, choosing the best analysis setup to capture the effects/phenomena characterizing its behavior may be non-trivial. Nonetheless, such a choice can influence or bias the analysis, potentially revealing or obscuring distinct types of information, depending on the underlying nature and physics of the system (Fig.~\ref{fig:fig1}F), as we are going to discuss in the subsequent analyses. 

In the case of aqueous systems, it is known that focusing solely on the first solvation shell can be limiting~\cite{crippa_detecting_2023,offei-danso_high-dimensional_2022}. It has been recently demonstrated how analyses extending beyond that (e.g., including up to the fifth nearest neighbor)\cite{martino2024data}, can offer a more comprehensive understanding of the system~\cite{donkor_machine-learning_2023, donkor_beyond_2024, foffi_correlated_2023, soper_is_2019, daidone_statistical_2023}. 
This raises important questions such as, e.g.: What is the effect of the spatial resolution on the results that can be effectively attained upon analysis? Is there an ``optimal resolution" to analyze a system? And is it possible to somehow determine it \textit{a priori}? Addressing such questions is  challenging, but at the same time particularly useful, especially in the study of systems about which little is known \textit{a priori}. 

Fig. \ref{fig:fig1} shows the results of analyses obtained calculating the LENS time series using a cutoff of $r_c=10$~\AA. While an $r_c$ around such value is commonly assumed to enable information-rich analyses of aqueous systems,~\cite{crippa_detecting_2023,martino2024data,caruso_timesoap_2023,donkor_machine-learning_2023,Capelli2022Ephemeral} we introduce an automated approach to systematically determine the optimal spatial resolution to study virtually any type of system. 
It is important to emphasize that the extractable information -- and the corresponding optimal cutoff -- depends both the internal physics of the system and the descriptor's effectiveness in capturing relevant physical features. The efficiency and signal-to-noise ratio are in fact intrinsic properties of each descriptor~\cite{martino2024data}. Here, using LENS as an example, we present a systematic, data-driven approach to determine the optimal resolution for studying the internal dynamic complexity and overall physics of the system. 

To identify key cutoff values in an agnostic manner, our approach begins with the radial distribution function $g(r)$ of the oxygen atoms, which describes the radial particle density around each molecule. As shown in Fig.~\ref{fig:fig2}A, the $g(r)$ plot exhibits sharp primary peak corresponding to the first solvation shell, followed by peaks representing the second, third, and subsequent shells. 
By locating the minima between these peaks, we identify characteristic spatial length scales inherent to the system under study. 
In this case, the key distances, marked by red circles in Fig.~\ref{fig:fig2}A, correspond to the values $r_c = \{3.3, 7.5, 8.5, 13, 10, 13, 15, 20, 30\}$~\AA. These values serve as meaningful cutoff radii for computing the descriptors. 
 
We calculated the LENS descriptor for all oxygen atoms throughout the trajectory, using the identified key distances as cutoffs. For each cutoff, our method applied Onion Clustering to analyze the time series and determine the number of statistically distinct environments detected. By gathering and comparing these results, we identified the ``optimal spatiotemporal resolution" for extracting physically relevant information from the trajectories. At this stage, the analysis depends solely on the chosen descriptor. 

Fig.~\ref{fig:fig2}B-G show the results obtained via Onion Clustering of the LENS time series computed with the different cutoffs. Fig.~\ref{fig:fig2}B illustrates the results of Onion Clustering applied to LENS time series computed with $r_c = 3.3$~\AA, corresponding to the first solvation shell. At all time resolutions, the analysis detected at most one dynamically persistent environment, the bulk (ice), while all other molecular data points were grouped into the unclassifiable data cluster (transitions). Even at the highest resolution, the fraction of unclassifiable data points was approximately 50\%, highlighting the limited effectiveness of considering only the first solvation shell for this aqueous system. This outcome is akin to conventional pattern recognition methods. Moreover, for time resolutions below $\Delta t>10$~ns, Onion Clustering fails to resolve any dynamically persistent clusters (Fig.~\ref{fig:fig2}B), as the fraction of unclassifiable data points (orange) reaches 100\%. These results show that restricting the analysis to the first solvation shell does not includes in the LENS descriptor enough information for a meaningful classification of the system's dynamics. 

Notably, when the second solvation shell is included using a cutoff of $r_c = 7.5$~\AA\ (Fig.~\ref{fig:fig2}C), Onion Clustering identifies 2 to 3 dynamically persistent environments. Crucially, the ice/water interface becomes discernible within this time series for resolutions between 2 and 
8~ns. This observation aligns with findings reported for other descriptors~\cite{martino2024data, lionello2024relevant}, which highlight the existence of a lower time resolution threshold necessary to identify certain environments. In fact, for $\Delta t>8$~ns, the analysis loses the ability to distinguish the ice/water interface as a separate environment.
Interestingly, Fig.~\ref{fig:fig2}C also shows that at very high time resolutions ($\Delta t < 2$~ns), information about the ice/water interface is again lost, reducing the number of detectable clusters to two. This, as recently discussed ~\cite{martino2024data, lionello2024relevant}, is due to the fact that there is a minimum observation time which is required so that transitions can be robustly distinguished from the more dominant states, and thus classified as separate environments. 

When the cutoff distance is increased beyond 8.5~\AA\ (Fig.~\ref{fig:fig2}D-G), including the third solvation shell, the LENS time series consistently resolve three dynamically persistent clusters down to a resolution of $\Delta t \sim 10$~ns. This clustering stability persisted up to a cutoff distance of approximately 15~\AA. In other words, spatial resolutions in the range of $8.5$~\AA to 15~\AA  offer the best performance, as they maximize the detection of distinct clusters (Fig.~\ref{fig:fig2}D-F) and minimize information loss (orange curve) for time resolutions $\Delta t<10$~ns. For this system, these spatial resolutions, when applied to the analysis of LENS time series, enable the extraction and representation of the system's dominant features with optimal efficiency. 

\begin{figure*}[ht!]
    \centering
    \includegraphics[width=.9\textwidth]{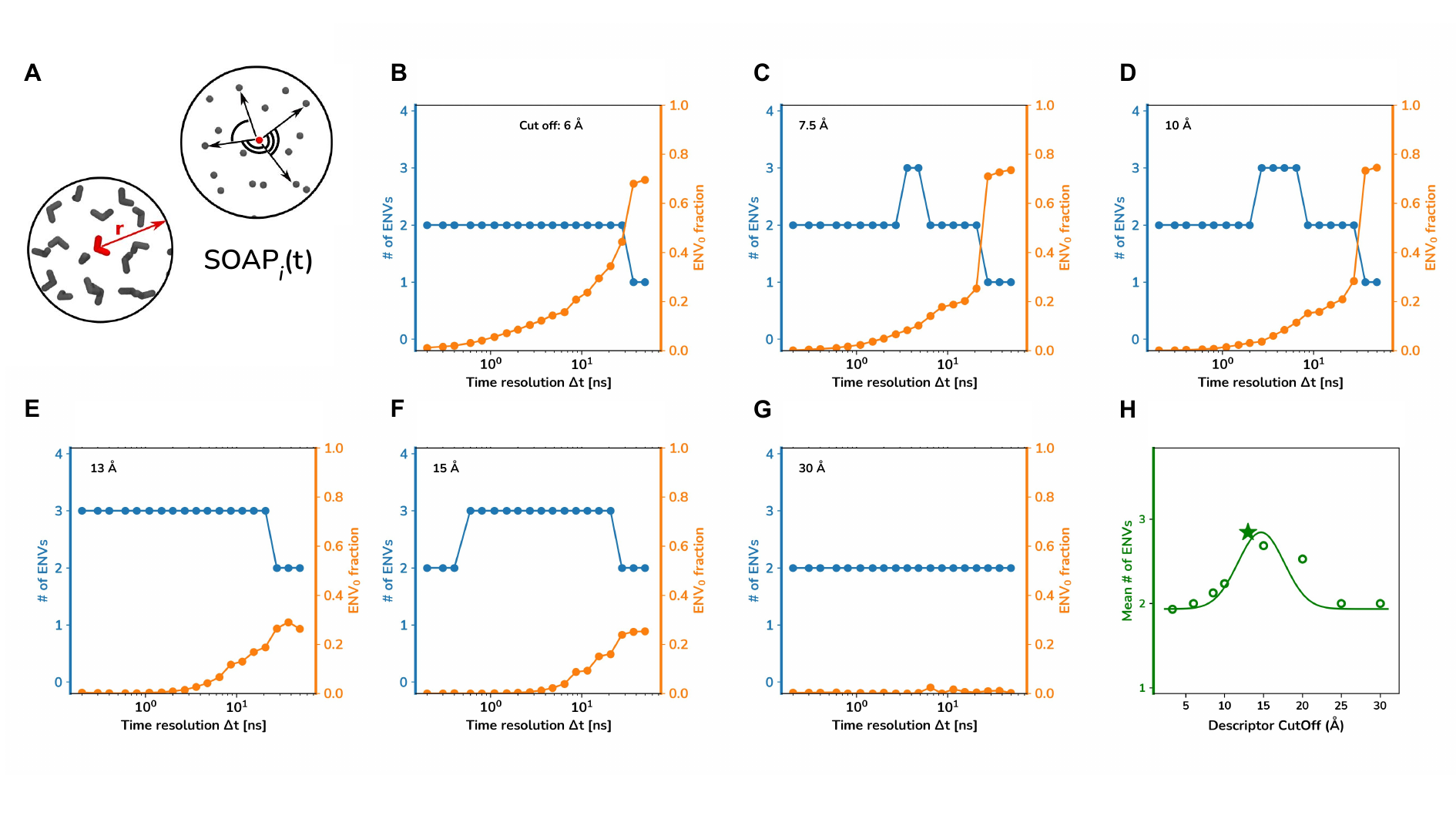}
    \caption{\textbf{Effect of spatial and temporal resolutions on information extraction from, \textit{e.g.}, SOAP time series data.} 
    \textbf{A} Schematic representation of SOAP. For each molecule $i$ in the system, at each sampled timestep $t$, the SOAP vector contains information on the distances and spatial displacements of the neighbors within a sphere of cutoff $r_c$: the SOAP power spectrum is thus a fingerprint of the local neighbors density around every SOAP center (the oxygen atoms of each water molecule, in this case).
    \textbf{B-G} Onion plots reporting the results of the clustering analyses of the SOAP PC1 time series as a function of the cutoff $r_c$ values used to calculate the SOAP power spectra. 
    \textbf{H} In green: mean number of classifiable micro-clusters (ENVs) before the fraction of unclassifiable data reaches 50\%, as a function of the $r_c$ used in the analysis. The data show a clear trend (green curve), where the maximum efficiency in information extraction-and-discretization is obtained at $13 \leq r_c \leq 15$~\AA, where the interface is detected in a robust way down to time-resolutions of $\Delta t<20$ ns. 
    }
    \label{fig:fig3}
\end{figure*}

When the LENS cutoff radius was set to values beyond $15$~\AA, the effectiveness of the analysis began to diminish. We extended the tests up to the maximum cutoff of 30\AA, constrained by the size of the simulation box. 
As shown in Fig.~\ref{fig:fig2}B-G and in supplementary Fig. S1, the ability to distinguish the ice/water interface progressively deteriorates with increasing cutoffs. At higher values, the analysis yielded results similar to those obtained with smaller cutoffs, where only the solid ice phase can be classified as a stable environment. 

For any trajectory, this method determines the optimal spatiotemporal resolution that maximizes the extraction of relevant information from noisy data. 
To summarize and compare results across different systems, we define a metric, $\mathcal{Q}$, which quantifies the quality of the clustering outcome. $\mathcal{Q}$ is calculated as the average number of stable environments identified by Onion Clustering across all time resolutions ($\Delta t$) where the fraction of unclassified data points is less than 50\%. 
Fig.~\ref{fig:fig2}H plots $\mathcal{Q}$ as a function of the cutoff radius used for LENS calculations. The plot reveals a clear peak in clustering quality for 1~nm~$<r_c<$~1.5~nm, highlighting this range as optimal for maximizing the information extracted from the analysis. With this cutoff radius, the maximum number of environments is identified using $\Delta t \leq 10$~ns. 

This result provides critical insights into the physics of the system, revealing the characteristic scales of the environments that dominate its dynamics. In this case, this ice/water system exhibits internal dynamics driven by collective events with specific spatial and temporal scales. For instance, the ice/water interface has a characteristic thickness of around 1.5~nm and contains molecules with a residence time of up to $\sim10$~ns (Fig.~\ref{fig:fig2}I). 
Clearly, this ``optimal spatiotemporal resolution" depends on the dynamics of each system (see next paragraph) and on the capability of different descriptors to capture them. Each descriptor emphasizes distinct features and has its own signal-to-noise ratio~\cite{martino2024data}. 
To illustrate the generality of these concepts, we repeated the Onion Clustering analyses on the first principal component (PC1) time series of the SOAP spectra computed for each molecule, calculated with different cutoff radii as done for LENS. This serves as another example of a descriptor providing rich information on the system (see Method for details). 

The results shown in Fig.~\ref{fig:fig3} show a very similar trend as those in Fig.~\ref{fig:fig2}. 
As in the LENS analysis, Fig.~\ref{fig:fig3}A-G demonstrates the existence of an optimal spatial resolution that maximizes the information extracted from the SOAP PC1 time series. Below this resolution, the analysis lacks sensitivity, and above it, the time series become too noisy to resolve the ice/water interface. All other explored spatial resolutions are shown in supplementary Fig. S2. Also MD representative snapshot of this system, all taken at $\Delta t = 40$~ns are shown in supplementary Fig. S5.  
The plot in Fig.~\ref{fig:fig3}H shows a trend similar to Fig.~\ref{fig:fig2}H, revealing an optimal spatial resolution at $r_c=13-15$~\AA. With this $r_c$, the interface is detectable using a time-resolution $\Delta t< 20$~ns (Fig.~\ref{fig:fig3}D-E). 
Interestingly, the slightly higher optimal cutoff radius for SOAP compared to LENS reflects differences between the descriptors: while LENS captures exchanges between molecules inside and outside the cutoff, SOAP considers only the local environment within the cutoff. This makes the SOAP optimal radius consistent with the detection of events at the fourth solvation shell, compared to the third shell for LENS, while capturing the same underlying physical phenomena. 
Let’s now apply these concepts to another system with completely different dynamical complexity to investigate how the approach performs in distinct scenarios. 

\begin{figure*}[ht!]
    \centering
    \includegraphics[width=.9\textwidth]{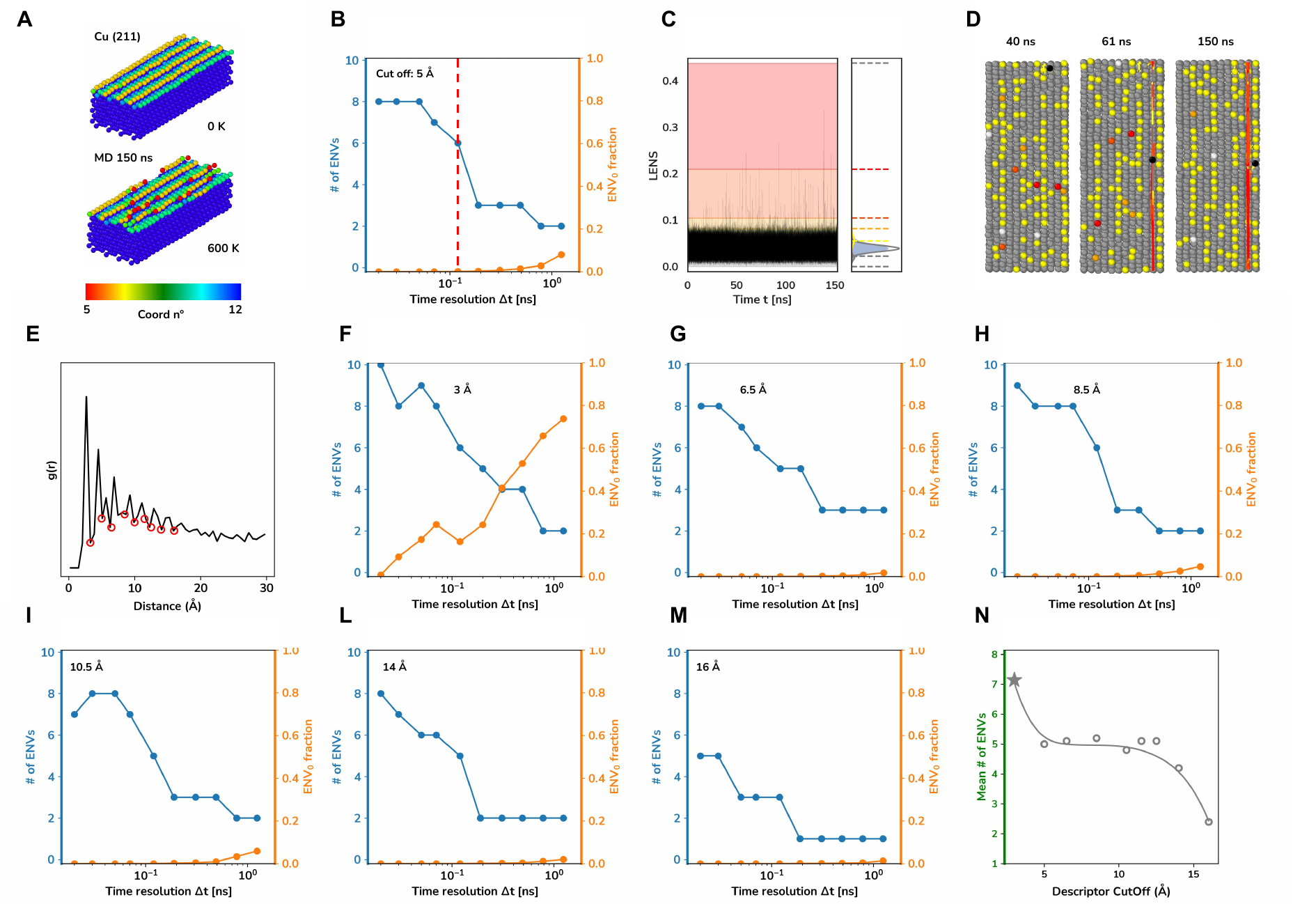}
    \caption{\textbf{Optimal resolution in the study of a dynamic metal surface.} 
    \textbf{A} MD snapshots of a Cu surface composed of 2400 atoms simulated at $T=600$~K: the atoms are colored according to their coordination number. Top: crystalline surface at $T = 0$~K before simulation start; bottom: representative snapshot of the surface equilibrated at $T = 600$~K. 
    \textbf{B} Onion plot: number of clusters (in blue) classified from the LENS time series obtained with cutoff radius $r_c = 5$~\AA, and fraction of unclassifiable data points (orange) as a function of the time-resolution $\Delta t$ used in the analysis. The vertical red dashed line indicates $\Delta t = 0.12$~ns as an example time-resolution used to plot the analysis results in panels \textbf{C-D}. 
    \textbf{C} LENS time series for all Cu atoms in the surface model along the entire MD simulation of $\tau=150$~ns (sampled every $\Delta\tau=10$~ps - see Methods for complete details). At the resolution of $\Delta t = 0.12$~ns, Onion Clustering classifies six different micro-clusters (colored areas) whose thresholds are identified by horizontal colored dashed lines. 
    \textbf{D} Top view of the simulation box at three different simulation times: $t=40$, $61$, and $150$~ns. An example Cu atom (ID = 144) is highlighted in black, whose history trajectory up to that point is colored according to the LENS micro-cluster it belonged to. At 150~ns, a red/orange straight vertical sliding on the (211) surface edge is clearly visible. 
    \textbf{E} Radial distribution function $g(r)$ of the Cu atoms in the system: relevant spatial cutoffs are highlighted by red circles (relevant $g(r)$ minima). 
    \textbf{F-M} Onion plots reporting the results of the analyses on LENS time series with different $r_c$. Same color coding as before. 
    \textbf{N} Mean number of clusters (ENVs) classifiable (before ENV0 reaches 50\%) as a function of the $r_c$ used in the analysis. In this case, dominated by local single-atom dynamical events, the optimal resolution is achieved at the smallest $r_c$: i.e., when accounting only for the first neighbors shell. }
    \label{fig:fig4}
\end{figure*} 

\subsection*{Optimal resolutions for a metal surface dominated by local dynamic events}
To demonstrate the generality of our approach, we applied it to analyze LENS time series computed from simulation of an atomistic model of a copper (211) surface simulated at 600~K (Fig.~\ref{fig:fig4}A) composed by 2400 Cu atoms replicating on the $xy$ plane via periodic boundary conditions, using a deep neural network potential trained to guarantee DFT-level accuracy (see Method section for details)~\cite{cioni_innate_2023}. As recently demonstrated, even well below the melting temperature $\sim1400$~K, such system demonstrates a peculiar dynamics, with sparse individual atoms migrating along the surface crests and diffusing sliding rapidly over the surfaces face, being then reincorporated into the crests~\cite{cioni_innate_2023, crippa_detecting_2023, crippa_machine_2023, caruso2024classification}. These rare local events are challenging to capture using conventional pattern recognition method, while they can be effectively captured \textit{via} Onion Clustering analyses of LENS time series~\cite{becchi_layer-by-layer_2024}. 

Using this MD trajectory, we followed the same procedure illustrated in the previous paragraph, computing LENS using different cutoff radii (using different $g(r)$ minima, Fig.~\ref{fig:fig4}E) and then measuring, via Onion Clustering, how much information can be extracted for each time series dataset. As shown in Fig.~\ref{fig:fig4}, Onion Clustering can detect different environments. For every choice of the cutoff radius, the clustering can identify various, more or less dynamic, environments in the system (Fig.~\ref{fig:fig4}B-C, for instance, using $r_c=5$~\AA) corresponding, e.g., to the metal bulk or static surface domains, as well as the crests and diffusing atoms on the surface (Fig.~\ref{fig:fig4}D). 
The trend in the number of resolvable clusters (Fig.~\ref{fig:fig4}F-M: in blue) shows how the number of distinct environments that Onion Clustering can resolve in the system decreases as the cutoff radius used in the analysis increases (Fig.~\ref{fig:fig3}B,F-M). In Fig.~\ref{fig:fig4}F, for example, it is evident that up to 8-10 clusters can be resolved using $r_c=3$~\AA (first neighbors shell) up to a temporal resolution of $\Delta t \sim 100$~ps (the faster dynamics occurring in this system require an higher temporal resolution than the ice/water system studied previously). As the cutoff increases, the number of clusters tends to decrease, reaching a maximum of 8 for a cutoff radius in the range $6.5 \leq r_c\leq10.5$~\AA. For $r_c>14$~\AA, there is a considerable drop in the number of resolved clusters, and above $r_c = 15$~\AA the effectiveness of the analysis deteriorates, and it ends up detecting only the metal bulk and fluctuations of the surface. All other explored spatial resolutions (red circles in Fig.~\ref{fig:fig4}E) are shown in supplementary Fig. S3. 

Fig.~\ref{fig:fig4}N sums up this monotonic drop in the clustering quality $\mathcal{Q}$ with increasing LENS cutoff radius, and reveals how, for this system, the optimal resolution is met at the shorter spatial length scales (3~\AA) and for time-resolution $\Delta t\leq100$~ps. This demonstrates how the physics of such system is completely different from that of the ice/water system, being dominated by local events (like the diffusion of individual Cu atoms along the surface). In this case, increasing the cutoff leads to information loss, due to the fact that the physics of the events characterizing the system is determined at short length scales, and nothing relevant can be effectively observed (by LENS) beyond the first neighbor shell. 
To conclude our study, we add to our comparison a completely different mesoscopic-scale system, with times-series data extracted from experimental data. 

\subsection*{An example experimental complex system dominated by collective events}

\begin{figure*}[htbp]
    \centering
    \includegraphics[width=0.9\textwidth]{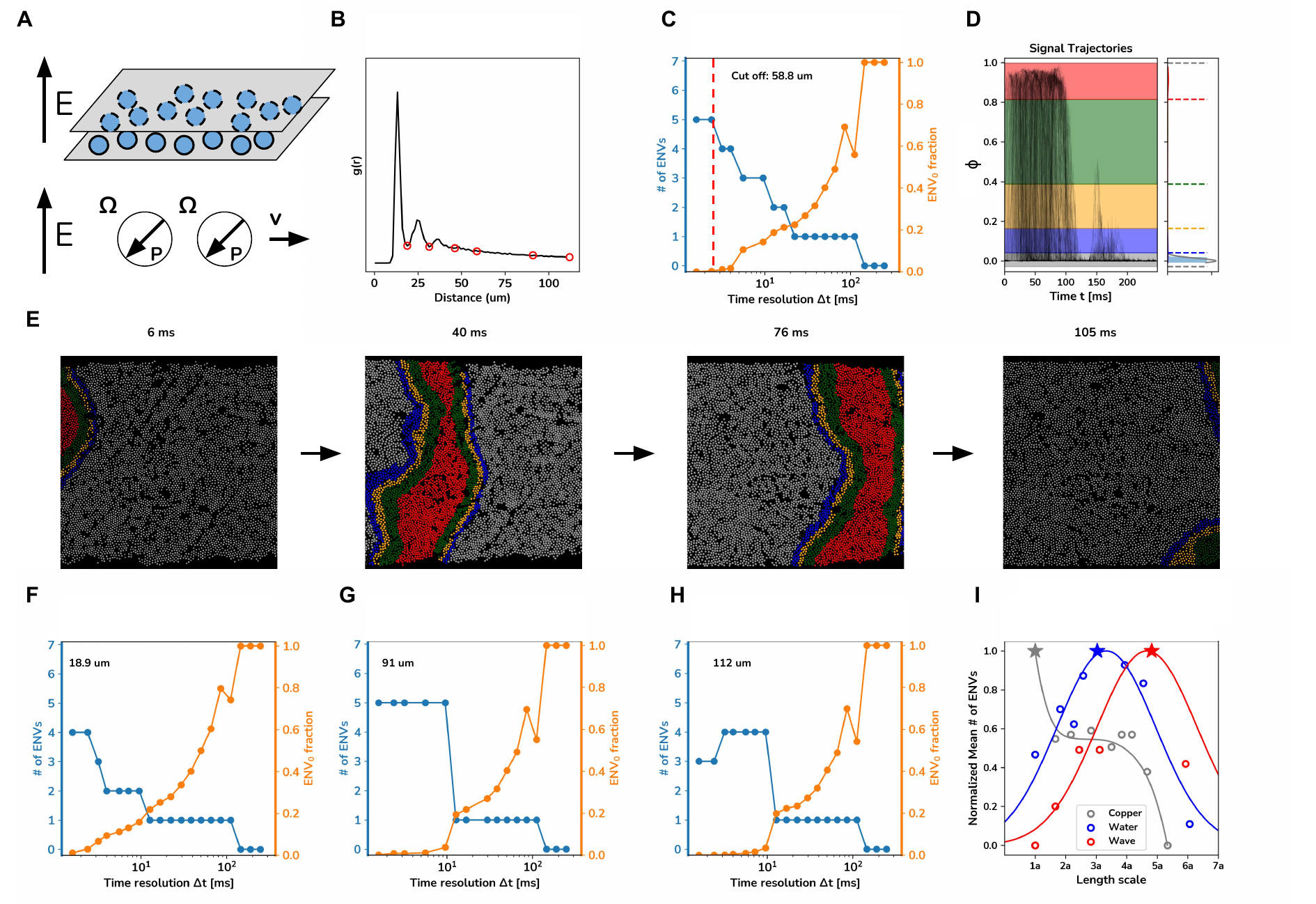}
    \caption{\textbf{Optimal resolution and characteristic length-scales in experimental complex systems: \textit{e.g.}, collective waves in Quincke rollers colloids.} 
    \textbf{A} Cartoon representation of Quincke rollers: dielectric colloidal microparticles confined in the $xy$ plane that, immersed in a conducting fluid, give rise to complex collective motions, waves, vortexes, etc., under exposure to a weak DC electric field (orthogonal to the $xy$ plane). We analyze trajectories resolved from an experimental microscopy movie taken from Ref.~\cite{liu_activity_2021}, where 6921 Quincke rollers undergo a collective wave that crosses a microscopy field of $700\times700$~$\mu$m$^2$ left-to-right along $\tau = 0.25$ s of observation. We measured the average local velocity alignment using a sphere of radius $r_c$ (Eq.~\ref{eq:eq1}). 
    \textbf{B} Radial distribution function $g(r)$ of the particles in the system, and characteristic $r_c$ distances identified by red circles. 
    \textbf{C} Onion plot for the analysis of the $\phi$ time series computed with cutoff radius $r_c = 58.8$~$\mu$m: number of clusters (in blue) and fraction of unclassifiable data (orange) as a function of the time-resolution $\Delta t$ (the red dashed line identifies the $\Delta t = 2$~ms time-resolution, for which results are shown as an example in panels \textbf{D-E}). 
    \textbf{D} The five clusters resolved by Onion Clustering at the example time-resolution, colored in gray, blue, yellow, green, and red (ordered with increasing $\phi$). In red is clearly visible the core of the wave crossing the microscopy field (see also Supplementary Movie S1). 
    \textbf{E} Four snapshots taken from the trajectory, where the Quincke rollers are colored according to the cluster they belong to. 
    \textbf{F-H} Onion results of the analysis of the $\phi$ time series calculated with different $r_c$ values. In this case, the maximum efficiency in information extraction-and-classification is encountered at $r_c=91$~$\mu$m, corresponding to the $\sim5th$ neighbors shell. 
    \textbf{I} Number of clusters resolved in each system, rescaled so that the maximum of each curve is equal to 1, as a function of the cutoff radius $r_c$ used for the computation of the descriptor. In order to compare the different systems, $r_c$ is expressed in multiples of the first neighbors shell radius $a$. }
    \label{fig:fig5}
\end{figure*}

As a final test case, we focused on a recently reported active matter system, composed by mesoscopic polystyrene particles, known as Quincke rollers. Under certain conditions, these particles exhibit highly non-trivial collective dynamics, such as emerging waves, vortexes, etc~\cite{liu_activity_2021} (Fig.~\ref{fig:fig5}A). 
For specific details about the system and how the analyzed trajectories were obtained, we refer the interested reader to Refs.~\cite{liu_activity_2021, becchi_layer-by-layer_2024}. In particular, the analyzed trajectory is reconstructed from a microscopy movie, with a microscopic field of $700\times700$~$\mu$m$^2$, containing 6921 polystyrene particles (with a diameter of $\sim10$~$\mu$m), which showed a wave span a large area of the system from left to right. The particles' positions in the $xy$ plain when exposed to a weak perpendicular electric field were tracked for a duration of approximately 0.25~s. 

Also for this mesoscopic system we performed an analysis similar to those presented in previous cases. Fig.~\ref{fig:fig5}B, shows the $g(r)$ computed from the trajectory, identifying the critical distances, marked by red circles (such distances characterize typical neighbors shells within the system, as seen previously). 
In this case, we employed a physic-inspired descriptor widely used for studying active matter ~\cite{ginelli_physics_2016,aung_local_2024}, local velocity alignment $\phi$. Also for this system, we denoised the descriptor using local averaging. The resulting quantity measures the average local velocity alignment in the environment surrounding each particle. This provides another demonstrative case of how our analysis is transferable to the study of time series obtained using any type of descriptor. 

The Onion Clustering analysis of the de-noised local velocity alignment time series detects several statistically distinct environments in this system. Fig.~\ref{fig:fig5}C-D shows the results produced by Onion Clustering using a cutoff of 59~$\mu$m (Including up to the third neighbor shell, Fig.~\ref{fig:fig5}B) with a time resolution of $1.6$~ms (vertical red dashed line, Fig.~\ref{fig:fig5}C). With this spatial resolution, five clusters can be distinguished. The gray cluster includes the majority of the particles, having $\Phi_i\sim 0$, which are not moving or with small random relative motions. The red cluster includes the particles with the highest values of $\Phi_i$, located in the inner part of the wave, moving with the maximum alignment in a coordinate fashion. The three remaining clusters (green, yellow, blue) correspond to intermediate alignments velocities area, from highest to lowest values respectively (Fig.~\ref{fig:fig5}D).
Looking at Fig.~\ref{fig:fig5}E, it is possible to refer such clusters to well-defined physical regions in the system ,where the red area corresponds to the center of the wave crossing left-to-right the microscopic field over the observation, while the blue,yellow and green areas represent the wave edges (see also supporting Movie1). 

The panels F-H of Fig.~\ref{fig:fig5} show the results of the analysis using different cutoff radii, starting from the minimum cutoff (including only the first neighbors shell, $19$~$\mu$m) to the maximum one ($\sim100$~$\mu$m, corresponding to the 6th neighbor shell). The minimum cutoff ends up to $r_c\sim19$~$\mu$m (i.e., accounting only for the first neighbor shell), thus the analysis resolves up to 4 clusters. When accounting for the fourth neighbor shell ($r_c\sim58.8$~$\mu$m), the analysis can better discretize the system, detecting up to 5 clusters (Fig.~\ref{fig:fig5}C). At $r_c\sim91$~$\mu$m cutoff, which includes up to the fifth $g(r)$ peak, the analysis resolves in a robust way up to 5 clusters for a temporal resolution in the range 1.6~ms~$\leq\Delta t\leq10$~ms (Fig.~\ref{fig:fig5}H). 

The value $r_c = 91$~$\mu$m, corresponding to the fifth neighbors shell, is found to be the optimal spatial resolution for resolving the internal complexity of this system. In fact, such characteristic length scale correspond to the width of the collective wave traversing the system (Fig.~\ref{fig:fig5}E:in red). 
Notably, using $r_c\geq91$~$\mu$m means using a cutoff larger that the wave thickness which results in reducing the efficiency of the analysis: using $r_c\sim112$~$\mu$m is found detrimental, and the analysis resolve again at most 4 distinct clusters (Fig.~\ref{fig:fig5}H). Essentially, this analysis reveals the characteristic size of the dynamic physical events captured by the descriptor within the system. 
All other explored spatial resolutions (red circles in Fig.~\ref{fig:fig5}B) are shown in supplementary Fig. S4. 

We point out that here the important point is not the exact number of clusters identified (which depends on the Onion Clustering analysis and the physical meaning of the descriptor). 
Rather, what matters is that at the same time resolution, a certain specific spatial resolution allows for extracting and better clustering in a statistically robust way the physics of the system: i.e., a purely data-driven assessment of the optimal resolution to attain as much relevant information as possible on the physics of such system from the data. 

Fig.~\ref{fig:fig5}I plots the analysis quality $\mathcal{Q}$ (the number of detected clusters, averaged over the time resolutions for which the unclassified data is less than 50\%) as a function of $r_c$. This plot reports the results for the three different systems studied in this work, rescaling the $x$-axis so that $r_c$ is expressed in terms of distances from the first, second, third neighbors shell etc. (i.e., a topological distance rather than a metric one). This allow us to compare systems with different length scales (atomic vs. molecular vs. mesoscopic). 

It is evident that these three cases correspond to three completely different behaviors, as it is captured by our analysis. The metal system, which is controlled by local events, shows the best resolution at short distance (first neighbor), while accounting for longer-range effect leads to information loss (gray curve). The ice/liquid water system is on the other hand controlled by more collective events that involve up to the third-fourth neighbors. Lastly, the active matter complex system composed of colloidal rollers has an internal physics dominated by collective system-wide events.
Such an approach can thus allow identifying directly from the data of the motions of the units that constitute a system the characteristic (spatiotemporal) length scales dominating its physical behavior, and the best resolutions to analyze them. Furthermore, its generality makes it suited to be applied to virtually any complex dynamical system (provided that a trajectory of the constitutive units can be attained). We believe that this approach will be helpful not only to study and understand the behaviors of systems about which little is known \textit{a priori}, but also to assist any type of data analysis suggesting the best resolutions to be used in space and time.

\section*{Conclusions}
Identifying the best resolutions to study a certain system in space and time might be far from trivial. Typical analyses, even advanced ones, often rely on physical intuition or prior experience and knowledge of the system under study, which fall short especially when dealing with systems about which little is known \textit{a priori}. 

On a practical level, this fundamental problem typically leads to very basic questions, such as: What is the effect of changing the cutoff in analyses conducted using distance-based local descriptors such as SOAP, LENS, coordination numbers, etc.? How to choose the optimal cutoff distance for the study of correlations, transitions, or fluctuations? What is the best inter-frame stride to choose in a time series analysis to ensure that key physical events are correctly captured? 
The approach that we report here offers an efficient way to answer such fundamental general questions. 

Dealing with the problem in a completely abstract way, our method does not build on any prior assumption and essentially learns the optimal space- and time-resolutions to study a determined system simply based on its data. 
In this way, the choice of, \textit{e.g.}, the best cutoff distance to be used to translate the system’s trajectories into information-rich time series data and the best time-interval ($\Delta t$) to extract such information from them, stop being parameters (to be tuned) in the analysis but are implicitly encoded in the physics and in the data of the system itself. 
The concept of “optimal spatiotemporal resolution” provides a fully unsupervised data-driven approach to identify the optimal spatiotemporal resolutions to characterize any type of complex dynamical system for which a trajectory of the constitutive units is resolvable. The results of the analysis remain thus dependent only on the type of descriptor(s) used (and the type of information that this/these may provide)~\cite{martino2024data} and on the resolution in the original acquisition of the raw data (\textit{i.e.}, the frame interval $\Delta \tau$), which sets a maximum in the time-resolution that it is possible to use in the study of the system (\textit{i.e.}, the limit is in the data themselves). 

Comparing the results obtained with two very different descriptors for the same system in Fig.~\ref{fig:fig2} \textit{vs.} Fig.~\ref{fig:fig3} (LENS \textit{vs.} SOAP), it is interesting to note that any specific descriptor translates the raw data in a different time series, which has its own optimal spatiotemporal resolution to be best studied. In particular, it is interesting to note that in both cases the optimal spatial resolution to study this aqueous system is found for $r_c$ values encompassing up to the 3$^{rd}$-4$^{th}$ neighbors' shells (descriptor dependent). At the same time, the optimal time-resolution is different: being a quite noisy local descriptor, SOAP suffers of information loss due to oversampling much more than LENS, which has a higher signal-to-noise ratio~\cite{martino2024data, caruso2024classification, lionello2024relevant, crippa_detecting_2023, becchi_layer-by-layer_2024}. 
While all analyses reported in Figs.~\ref{fig:fig1}-\ref{fig:fig5} refer to systems for which it has been possible to resolve the individual trajectories of the constitutive units composing them, it is worth noting that this is not always easy or possible. Nonetheless it is also worth noting that, given the generality of such optimal spatiotemporal resolution concept, this may offer a robust basis to identify optimal resolutions to study and extract information also from different types of data, such as, \textit{e.g.}, static data distributions, or time series data not derived from many-body systems: in such cases, the broader concept of optimal spatiotemporal resolution simply reduces respectively to that of optimal spatial or temporal resolution. 

While, from a purely technical point of view, this work offers a robust and transferable concept to support the extraction of information from the analysis of virtually any type of data (independently on whether this comes from simulations or experiments, or on the scale of the system under study), the physical information contained in it is at least as much important. 
As seen in Figs.~\ref{fig:fig2}-\ref{fig:fig5}, the analysis identifies different optimal spatiotemporal resolutions that essentially depend on the local or collective nature of the key microscopic-level events/processes dominating the various types of systems. Zooming locally in the study of the copper surface of Fig.~\ref{fig:fig4} allows extracting more information on the physics of this system -- which is characterized by the local motions of individual atoms over time --, while zooming-out lowers the resolution and leads in this case to information loss (Fig.~\ref{fig:fig4}N). \textit{Vice versa}, zooming too much locally provides an incomplete picture of the physics of a system dominated by collective events, which are better captured at larger $r_c$ distances (Fig.~\ref{fig:fig5}I: this is the case of, \textit{e.g.}, the Quincke rollers’ and of the ice/water systems). 

In general, being the optimal resolution for studying a certain system the one allowing to extract as much information as possible on its internal physics, this automatically unveils the nature of the physical events characterizing the system itself. This is a general principle that is particularly relevant for exploring in unbiased way and for understanding systems whose physics is unknown \textit{a priori} based on their observation, which is one of the cornerstones of physics. 
Such a data-driven assessment of ``optimal spatiotemporal resolutions" concept provides a simple, yet general and efficient approach for studying and understanding a variety of systems simply based on their data, as well as a practical approach to guide data analysis in general. 

\section*{Methods}
\subsection*{Water/ice coexistence system}

The data shown in Figs~\ref{fig:fig1}-\ref{fig:fig3} were obtained from a $50$~ns long MD trajectory of an atomistic model system containing 2048 TIP4P/ICE water molecules~\cite{crippa_machine_2023}. The system was simulated at the pressure of 1~atm, starting from a condition in which half of the molecules are in the liquid state and half were arranged in an hexagonal ice packing  in correspondence of the melting temperature ($T = 268$~K for this force field~\cite{conde_high_2017}). Further details on this system can be found in Refs.~\cite{crippa_machine_2023, caruso_timesoap_2023, crippa_detecting_2023}. The dynamics of the system were sampled for $ \tau = 50$ ns of MD simulation with a sampling interval of $\Delta \tau=0.1$ ns. 

\subsection*{LENS and SOAP time series}

For each one of the 2048 oxygen atoms, the LENS signals~\cite{crippa_machine_2023, crippa_detecting_2023} were computed on the sampled configurations. The different cutoff radii found from the $g(r)$ minima are used to compute different LENS time series, one for each cutoff ($r_c = \{3.3, 6, 7.5, 8.5, 10, 13, 15, 20, 30\}$~\AA). 

To describe the structural environment surrounding each particle within the simulations, we used SOAP. We compute the SOAP spectrum, representing the local structural environment of each particle at every timestep within a certain cutoff radius. The SOAP vectors are generated using dynsight platform~\cite{noauthor_dynsight_nodate}, setting the number of spherical harmonics and number of radial basis functions $n_\text{max} = n'_\text{max} = l_\text{max} = 8$. The results is a 576-component vector represent the environments of each particle at each timestep. Then, we applied the PCA algorithm to each dataset reducing the dimensionality to the first principal component. 
SOAP data series have been compute on the system considering the same oxygen atoms and the same cutoff radii as done for LENS. 

\subsection*{Onion clustering time series analyses}

For a detailed description of the Onion Clustering method, we refer the interested readers/users to the dedicated paper~\cite{becchi_layer-by-layer_2024}. All Onion Clustering analyses conducted in this work used the latest version of the algorithm as implemented and openly available in the dynsight platform~\cite{noauthor_dynsight_nodate}. 

In brief, Onion Clustering is an algorithm for single-point clustering of time series data, which allows to uncover and discriminate statistically relevant fluctuations from noise and to cluster them based on their similarity~\cite{becchi_layer-by-layer_2024, martino2024data, lionello2024relevant, caruso2024classification}. 
Considering a noisy time series of total length $\tau$ composed of $n$ time-intervals $\Delta \tau$, Onion Clustering performs a single point clustering of time series intervals of length set by the time resolution $\Delta t$ used in the analysis. In particular, while the domains classifiable in the time series depend on the time resolution used, instead of choosing $\Delta t$ \textit{a priori} Onion Clustering repeats iteratively the analysis from the maximum possible resolution ($\Delta t = 2 \times \Delta \tau$ in this method) to the minimum one ($\Delta t = \tau$) plotting the obtained results. Looking at the Kernel Density Estimation of the time series data, the Onion method detects the highest peak corresponding to a first dominant environment, fits a Gaussian distribution to it (\textit{i.e.}, an environment is characterized by the average value of the descriptor, and its variance), and counts the data intervals that remain in that environment for at least $\Delta t$. These data are then classified as part of that environment (stable at that time resolution) and their signals are removed from the time series that, cleared from their noise, is then analyzed again in search of new high density peaks. This allows to iteratively reduce the noise of the already-classified-data in the time series and to uncover, layer-by-layer, all microscopic domains that can be statistically classified in robust way at a certain resolution, including those hidden by the noise of the dominant ones. The method iterates until no new statistically relevant clusters can be uncovered. 
The fraction of data that cannot be classified at every $\Delta t$ is classified as part of the ENV0 cluster (i.e., transitions occurring faster than the time resolution of the analysis). 
In all the cases studied in this work, Onion Clustering analyses are conducted on the time series of all particles in the systems. 

As shown in the Onion plots of Figs.~\ref{fig:fig1}-\ref{fig:fig5}, performing such analysis at all values of the time resolution $\Delta t$ allows to automatically identify the optimal choice of $\Delta t$ that maximizes the discretization of the data into clusters and minimizes the fraction of unclassified points. The $\mathcal{Q}$ values in the plots of Fig. \ref{fig:fig2}H, \ref{fig:fig3}H, \ref{fig:fig4}N, \ref{fig:fig5}I are computed as the average of the number of resolved micro-clusters before the ENV0 population (\textit{i.e.}, the fraction of unclassifiable data) reaches 50\%.

\subsection*{Copper surface system analysis}

The data shown in Fig.~\ref{fig:fig4} were obtained from MD simulation described in detail in~\cite{cioni_innate_2023, becchi_layer-by-layer_2024, crippa_detecting_2023}, of an atomistic model of FCC(211) copper metal parametrized with a deep neural network potential~\cite{cioni_innate_2023}. The surface model contains 2400 Cu atoms that were simulated at $T = 600$~K. Periodic boundary conditions are applied in the $xy$ plane, while the system was finite along the $z$ direction, simulating \textit{de facto} an infinite copper/vacuum surface along the (211)-plane. The system is simulated for $\tau = 150$ ns, sampling its configuration every $\Delta \tau = 10$ ps, for a total of 15000~frames. LENS signals~\cite{crippa_detecting_2023} were then computed for all atoms. The descriptor time series have been generated for all cutoff in the list $r_c = \{3, 5, 6.5, 8.5, 10.5, 11.5, 12.5, 14, 16\}$~\AA. The signals were then smoothed \textit{via} moving average using a 10~frames window to reduce noise.\cite{crippa_detecting_2023,becchi_layer-by-layer_2024,caruso2024classification} 

\subsection*{Quincke rollers system analysis}
The data shown in Fig.~\ref{fig:fig5} were obtained by analyzing the trajectories of 6921 Quincke rollers obtained from an experimental microscopy video (reported in Refs.~\cite{liu_activity_2021})with a duration of $0.25s$. From the video, the $x$- and $y$-coordinates of 6921 particles for 310 consecutive frames are extracted, using in-house code~\cite{becchi_layer-by-layer_2024} and the tracking algorithm Trackpy~\cite{allan_trackpy_2016}, resulting in a $\Delta \tau=$0.8 ms. 

For each particle, we computed the local velocity alignment as 
\begin{equation}
    \phi_i \equiv \frac{1}{n_c^i} \sum_{j}^{n_c^i} \frac{\vec{v}_i\cdot\vec{v}_j}{|\vec{v}_i| |\vec{v}_j|}
    \label{eq:eq1}
\end{equation}
In Eq.~\ref{eq:eq1}, the index $j$ runs over the $n_c^i$ particles located within a specified cutoff distance from particle $i$, and $\vec{v}_i$ and $\vec{v}_j$ are the velocities of particles $i$ and $j$, respectively. Parameter $\phi_i$ ranges from -1 to 1: a value of 1 indicates perfect alignment, meaning particle $i$ and its neighbor particles within $r_c$ are moving in the same direction; a value of -1 indicates that velocities are in opposite directions, a value around 0 suggests that the particles are moving in perpendicular directions or exhibit no clear alignment (random relative motion). 
Denoising was then applied, assigning to each particle at each frame the the value $\Phi_i$, equal to the average values of $\phi_j$ within the sphere of neighbors. 
The time series were calculate for all cutoff in the list  $r_c = \{18.9, 31.5, 46.2, 58.8, 91, 112\}$~$\mu$m.

\begin{acknowledgments}
G.M.P. acknowledges the funding received by the European Research Council under the European Union’s Horizon 2020 research and innovation program (grant agreement no. 818776–DYNAPOL). This work has been also partly supported by FAIR - Future Artificial Intelligence Research and by the European Union Next-GenerationEU (Piano Nazionale di Ripresa e Resilienza (PNRR) – mission 4, component 2, investment 1.3 – D.D. 1555 11/10/2022, PE00000013, CUP: E13C22001800001). The authors also thank Dr. Chiara Lionello for the insightful discussions. 
\end{acknowledgments}

\section*{Data Availability Statement}

The data that support the findings of this study are openly available in \href{https://github.com/GMPavanLab/Optimal-Spatiotemporal-Resolutions}{this Github repository} (this link will be replaced with a definitive Zenodo link upon acceptance of the final version of this paper). 

\section*{Author contributions statement}
GMP conceived and supervised this work. DD, SM
and MB developed and performed the analyses. All
authors discussed the results. DD and GMP wrote the
paper. All authors contributed to the revision of the
paper.

\section*{Conflict of Interest Statement}
The authors have no conflicts to disclose.

\bibliography{paper.bib}

\clearpage

\setcounter{section}{0}
\setcounter{subsection}{0}
\setcounter{figure}{0}
\renewcommand{\thefigure}{S\arabic{figure}}
\onecolumngrid
\section*{Supporting Information}
\subsection*{Supporting text}
\label{SI:text}
\subsubsection*{Cleaning low populated micro states before evaluating clustering quality}
Once the time series data were calculated, we verified the robustness of the identified clusters prior to evaluating their quality. Specifically, any clusters with a population representing approximately less than $0.001$ of the total number of data points were deemed insufficiently robust and thus either merged or disregarded. Although we set $0.001$ as a general guideline, this threshold may be adjusted based on dataset-specific considerations. Further details regarding these parameter selections and adjustment criteria are available within the code implementation.

\subsection*{Movie S1}
Reconstruction of the experimental video used for the analysis of Fig5 in the main text, colored according to the clustering with $\Delta t$ = 1.6 ms. The 5 clusters found are: the red one correspond to the maximum $\Phi$ (wave's core), the gray one to elements which shown random mutual alignment during their movement ($\Phi\simeq$  0), the intermediate clusters (blue,yellow,green) going from lower to higher values of $\Phi$.

\clearpage

\subsection*{Supporting Figures}

\begin{figure*}[ht!]
\centering
\includegraphics[width=0.95\textwidth]{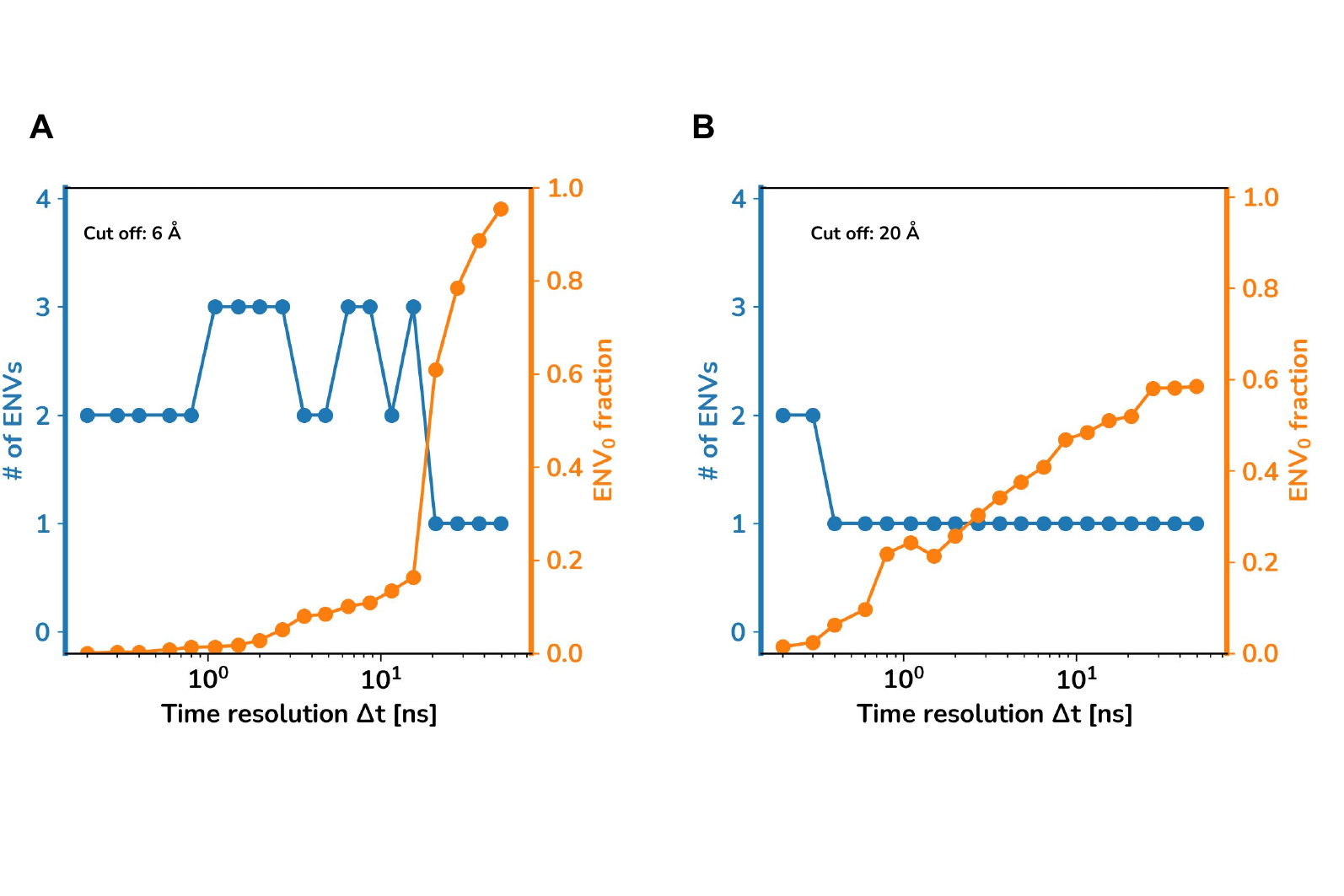}
\caption{\textbf{Supplementary cutoff of Fig 2 from main text.} 
    \textbf{A-B}: Onion plots obtained from the analyses of the LENS time series of water/ice system with $r_c=6$~\AA (A), $r_c=20$~\AA (B). The number of resolved micro-clusters is shown in blue, while the fraction of unclassifiable data (in the ENV0 cluster) is shown in orange.}
\end{figure*}
\clearpage

\begin{figure*}
\centering
\includegraphics[width=\textwidth]{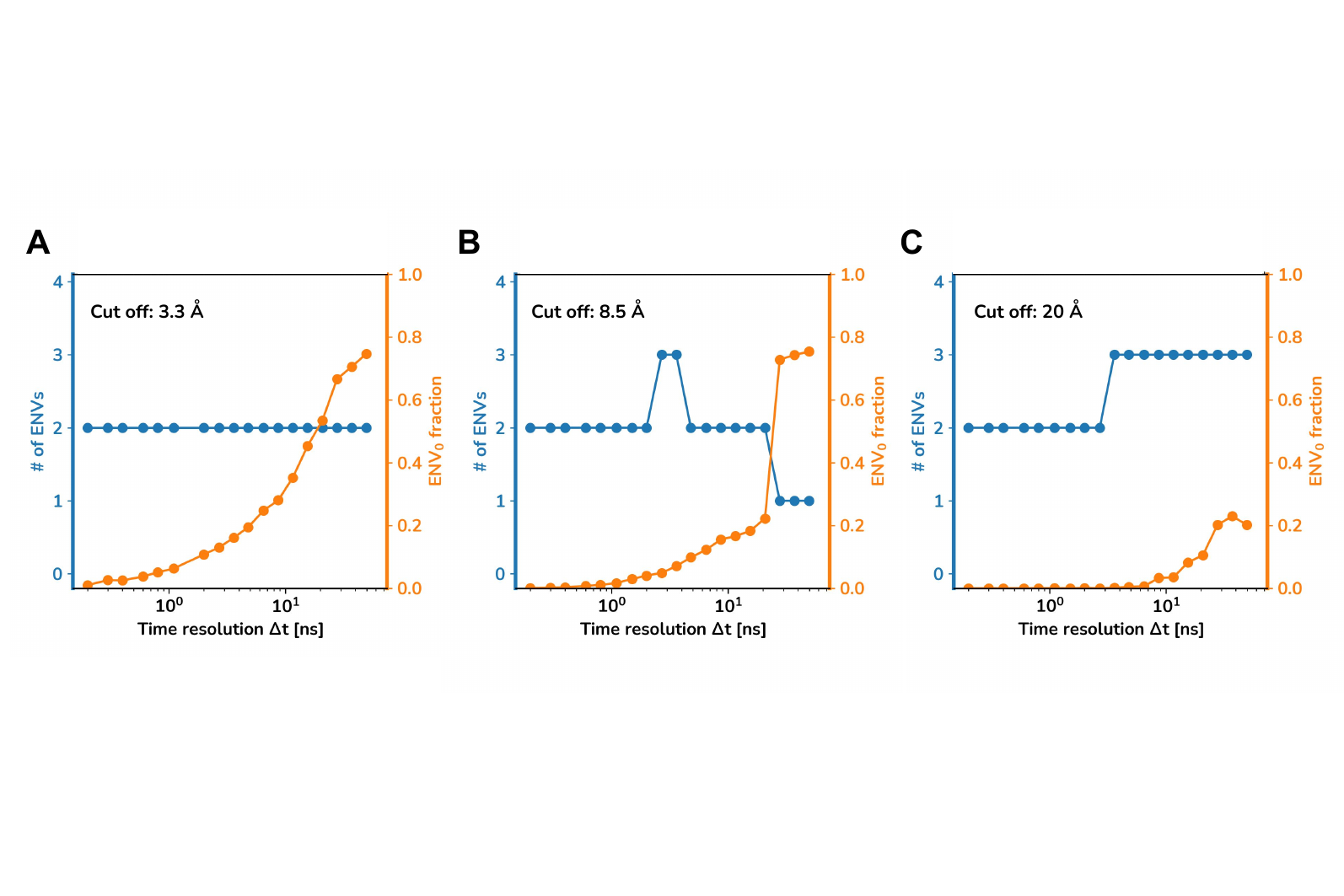}
\caption{\textbf{Supplementary cutoff of Fig 3 from main text.} 
    \textbf{A-C}: Onion plots obtained from the analyses of the SOAP time series of water/ice system with $r_c=6$~\AA (A), $r_c=8.5$~\AA (B), $r_c=20$~\AA (C). The number of resolved micro-clusters is shown in blue, while the fraction of unclassifiable data (in the ENV0 cluster) is shown in orange.}
\end{figure*}

\begin{figure*}
\centering
\includegraphics[width=\textwidth]{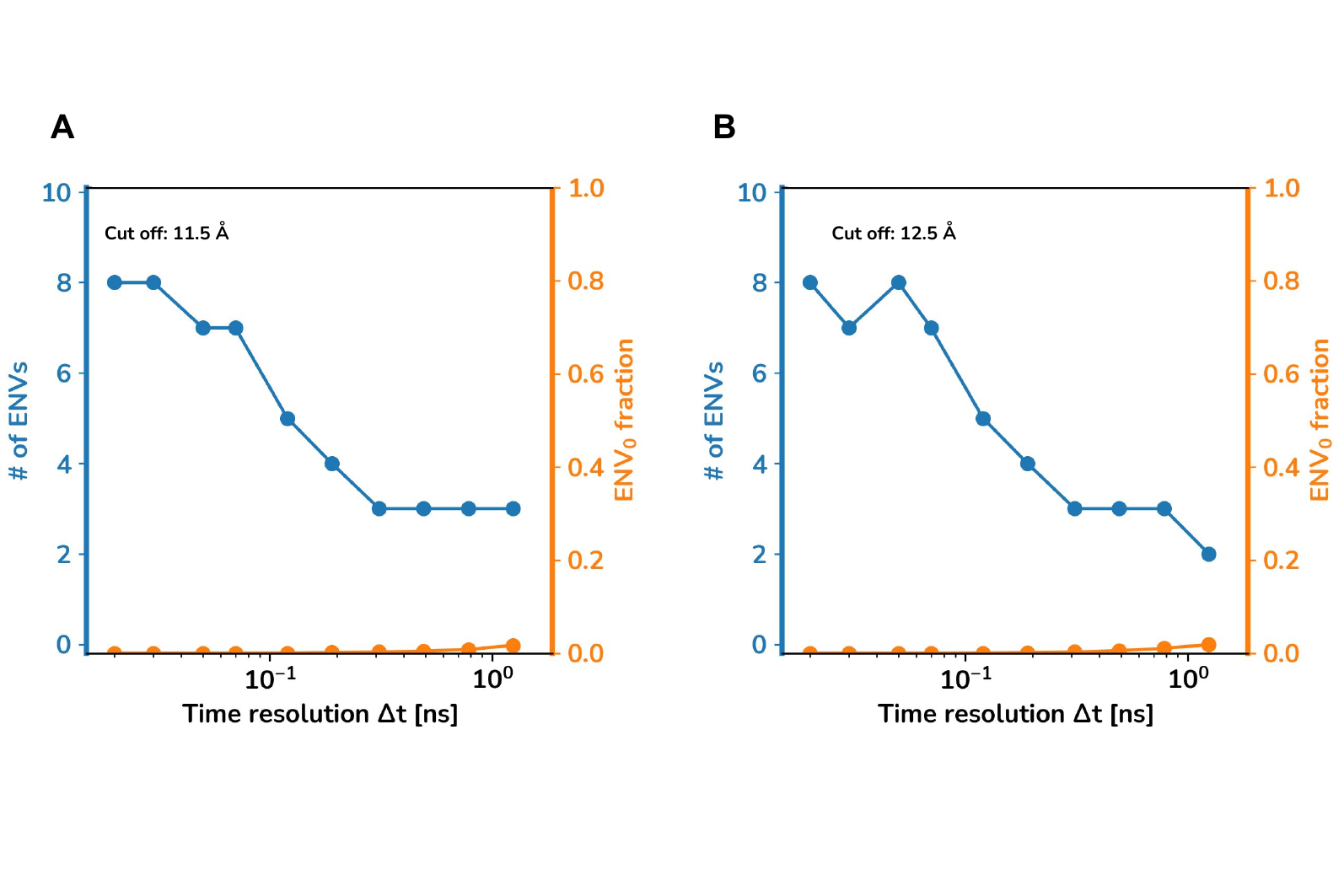}
\caption{\textbf{Supplementary cutoff of Fig 4 from main text.} 
    \textbf{A-B}: Onion plots obtained from the analyses of the LENS time series of copper system with $r_c=11.5$~\AA (A), $r_c=12.5$~\AA (B). The number of resolved micro-clusters is shown in blue, while the fraction of unclassifiable data (in the ENV0 cluster) is shown in orange.}
\end{figure*}

\begin{figure*}
\centering
\includegraphics[width=\textwidth]{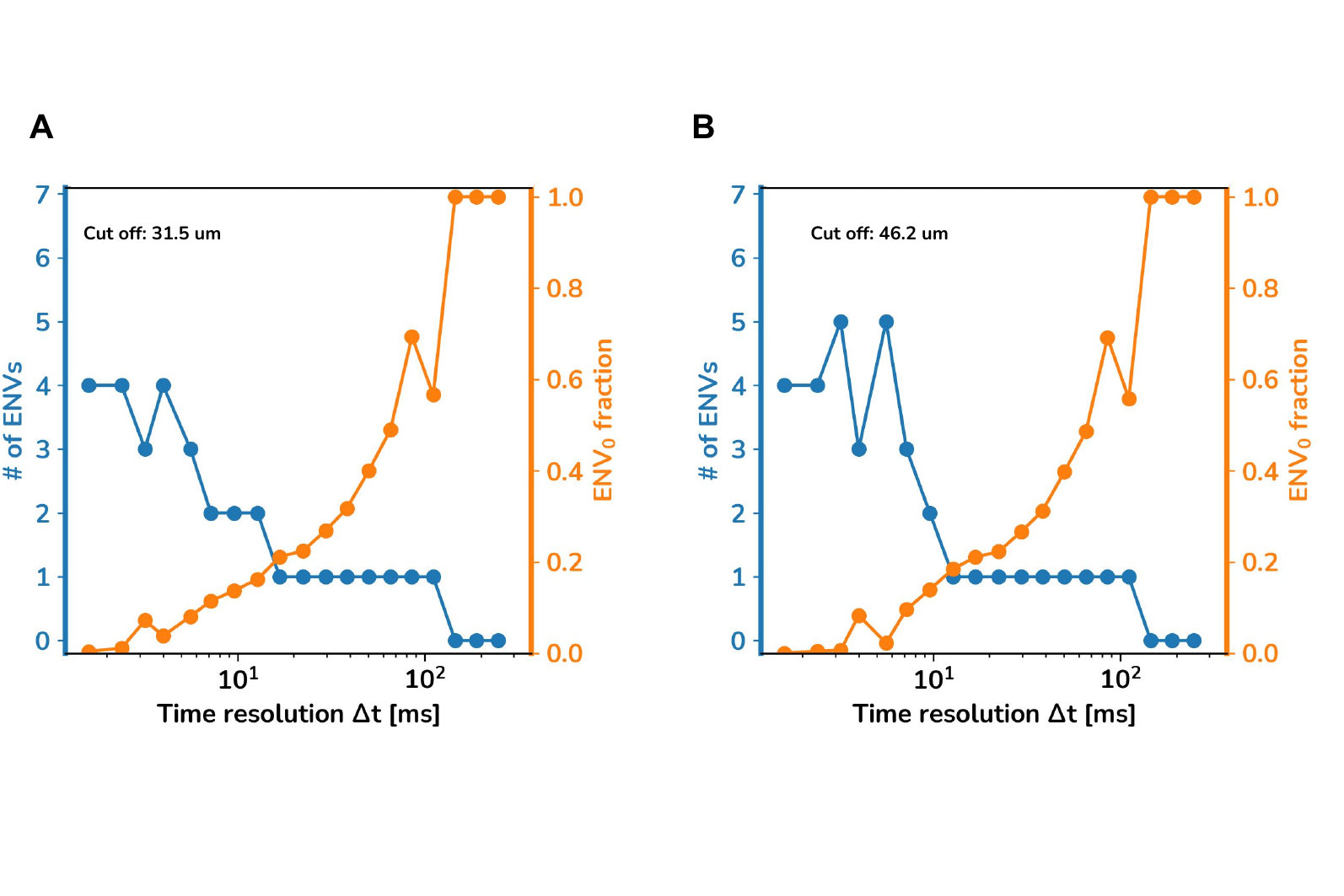}
\caption{\textbf{Supplementary cutoff of Fig 5 from main text.} 
    \textbf{A-B}: Onion plots obtained from the analyses of the $\Phi$ time series of Quincke rollers system with $r_c=31.5$~$\mu$m (A), $r_c=46.2$~$\mu$m (B). The number of resolved micro-clusters is shown in blue, while the fraction of unclassifiable data (in the ENV0 cluster) is shown in orange.  }
\end{figure*}

\begin{figure*}
\centering
\includegraphics[width=\textwidth]{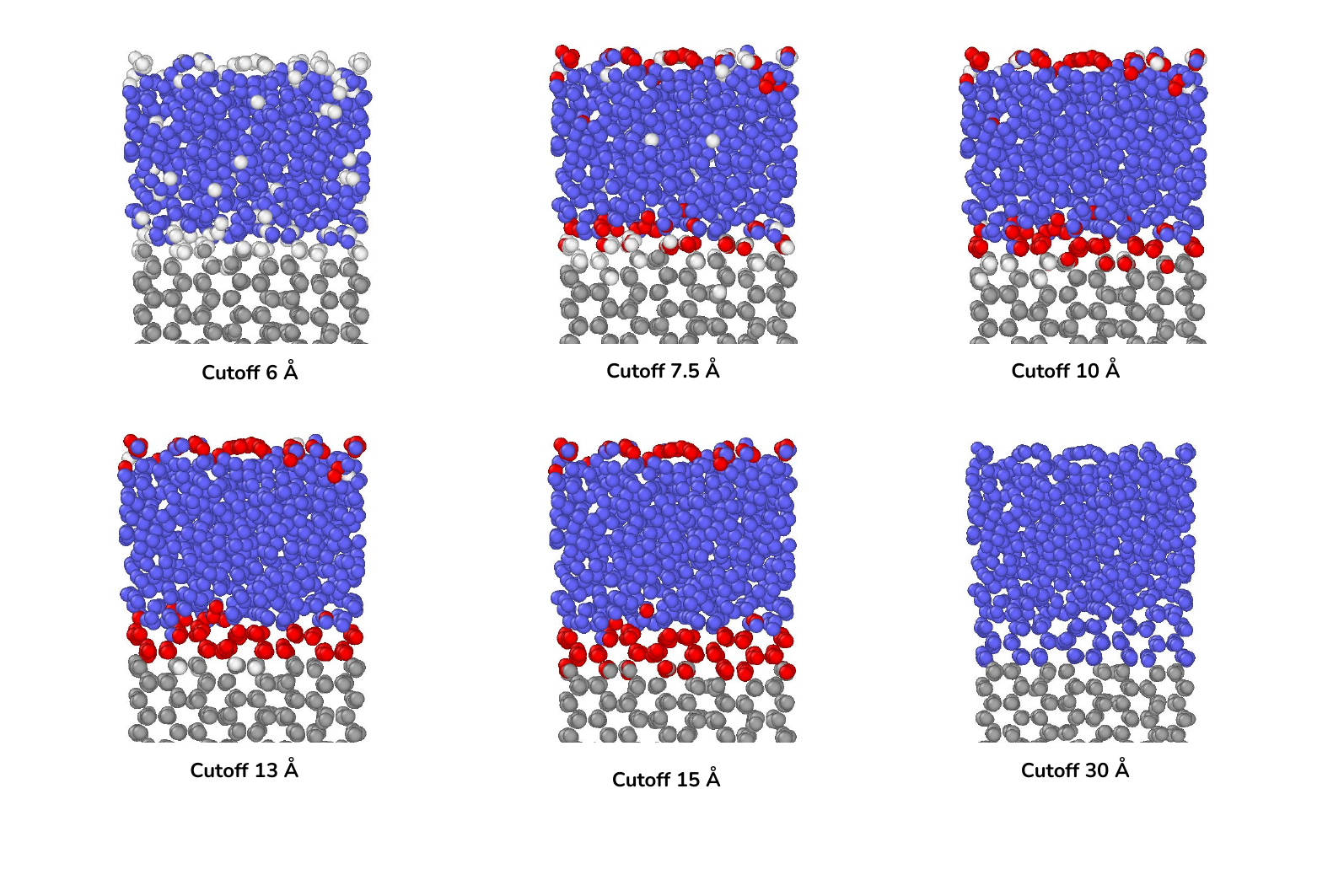}
\caption{\textbf{Snapshot of water/ice system.} 
    \textbf{} Representative MD snapshot of the water/ice system where the water molecules are colored according to the clustering obtained with the different $r_c$ in the analyses of panels \textbf{B-G} of Fig 3 from main text at the example time-resolution of $\Delta t=40$ ns.}
\end{figure*}

\end{document}